\documentclass[12pt]{article}

\usepackage[T1]{fontenc}
\usepackage[utf8]{inputenc}
\usepackage{lmodern}
\usepackage{microtype}
\usepackage[margin=2.5cm]{geometry}
\usepackage{setspace}
\usepackage{amsmath,amssymb,amsfonts}
\usepackage{bm}          % bold math
\usepackage{upgreek}     % upright Greek letters if needed
\usepackage{siunitx}     % SI units
\usepackage{graphicx}
\usepackage{float}
\usepackage{caption}
\usepackage[hidelinks]{hyperref}
\usepackage{url}
\usepackage{cleveref}
\usepackage{soul}
\usepackage{xcolor}

\usepackage[numbers,sort&compress]{natbib}
\newcommand{\Beff}{B_{\mathrm{eff}}}
\newcommand{\EF}{E_{\mathrm{F}}}
\newcommand{\vF}{v_{\mathrm{F}}}
\newcommand{\sigmaxy}{\sigma_{xy}}

\newcommand{\DeltaT}{\Delta T}
\newcommand{\Deltatheta}{\Delta\theta}
\newcommand{\DeltaF}{\Delta\theta_{\mathrm{F}}}
\newcommand{\thetaF}{\theta_{\mathrm{F}}}
\newcommand{\Deltasigmaxy}{\Delta\sigmaxy}

\title{\textbf{Light-induced effective magnetic fields in Landau quantized graphene}}

\author{
  Hiroki Ueda$^{1,*}$,
  Alexej Pashkin$^{2}$,
  Ece Uykur$^{2}$,
  Kryštof Kašša$^{3,4}$,
  Filip Chudoba$^{3}$,\\
  Jan Kunc$^{5}$,
  Manfred Helm$^{2}$,
  Milan Orlita$^{3,4}$,
  and Stephan Winnerl$^{2,*}$
}

\date{}

\begin{document}

\maketitle
\noindent
$^{1}$PSI Center for Photon Science, Paul Scherrer Institute, Forschungsstrasse 111, 5232 Villigen-PSI, Switzerland.\\
$^{2}$Institute of Ion Beam Physics and Materials Research, Helmholtz-Zentrum Dresden-Rossendorf, Dresden 01328, Germany.\\
$^{3}$LNCMI-EMFL, CNRS UPR3228, Univ.\ Grenoble Alpes, Grenoble 38042, France.\\
$^{4}$Department of Experimental Physics, Faculty of Science, Palacký University Olomouc, 17. listopadu 1192/12, 779 00 Olomouc, Czech Republic.\\
$^{5}$Faculty of Mathematics and Physics, Charles University, Ke Karlovu 5, Prague 121\,16, Czech Republic.\\[6pt]
$^{*}$Correspondence: \href{mailto:hiroki.ueda@psi.ch}{hiroki.ueda@psi.ch} and \href{mailto:s.winnerl@hzdr.de}{s.winnerl@hzdr.de}

\vspace{1cm}

% ============================================================
\section*{Abstract}
% ============================================================
Ultrafast magnetism triggered by circularly polarized radiation underpins ultrafast spin control, relevant to future technologies, e.g., opto-spintronics and magnonics.
The dynamics are often complicated and intertwined among correlated subsystems, such as electrons, spins, phonons, plasmons, topology, and lattice, due to many-body quantum coupling at ultrafast timescales.
Here, we demonstrate light-induced effective magnetic fields generated by selective excitation between non-equidistant Landau quantized states in graphene, a prototypical Dirac material, using circularly polarized pulses.
By magnetically tuning the Landau-level transition resonance away from other low-energy excitations, we obtain a clean electrostatically controllable platform and identify the microscopic origin of the light-induced magnetic signals, independent of sublattice coupling.
Because different Landau levels carry distinct optical Hall conductivities, direct modification of their occupancies via optical excitations creates transient Faraday rotation signals with dispersive magnetic-field dependence, mirroring the static magneto-optical lineshape.
The induced effective magnetic field normalized by the pump electric field exceeds typical reported  values for the inverse Faraday effect of electronic origin.
Our results establish a clear microscopic picture of the inverse Faraday effect of electronic origin, which can trigger hierarchical dynamics among correlated sublattices once Landau-level transitions are magnetically tuned to coincide with other low-energy excitations in Dirac systems and related materials.

\vspace{1cm}

% ============================================================
\section*{Main Text}
% ============================================================
\subsection*{Introduction}

Light-induced magnetic fields generated by circularly polarized electromagnetic radiation are central to the field of ultrafast magnetism~\cite{Kirilyuk2010}, in which laser pulses manipulate magnetic properties on ultrafast timescales. 
Representative phenomena include magnetization switching~\cite{Stanciu2007,Lambert2014}, magnon creation~\cite{Kimel2005,Satoh2010}, spin-polarized currents~\cite{Ganichev2002,Mclver2012}, and magnetic phase transitions~\cite{Sheu2019}, collectively underpinning opto-spintronics~\cite{Nemec2018}.
These magnetic fields emerge from non-thermal light-matter interactions that are phenomenologically described by the inverse Faraday effect, mediated by virtual or real electronic transitions~\cite{Pitaevskii1961,Battiato2014}, plasmon resonances~\cite{Cheng2020,Han2023}, or axial phonon resonances~\cite{Luo2023,Basini2024}.
The plasmonic mechanism, reported also in patterned graphene~\cite{Han2023}, is intuitive and essentially classical and exploits near-field enhancement using the extrinsic fabrication parameters, corresponding to magnetic field induction via a macroscopic circular current driven by the electric-field component of the radiation.
However, the remaining cases may involve intrinsic many-body quantum interactions among electrons, spins, and phonons, and are considerably less well understood. Identification of the primary channel that triggers ultrafast magnetism is essential to understand hierarchical dynamics and design light-induced emergent phenomena.

Two-dimensional Dirac materials, whose continuous energy spectrum with linear dispersion collapses into non-equidistant Landau quantized levels under a magnetic field $B$, offer an ideal platform for investigating light-induced magnetic fields via selective resonant excitation between individual Landau levels.
In contrast to conventional Schrödinger two-dimensional electron gas systems, where Landau-level energy is linear in the level index $n$ and $B$, the Landau-level energy in Dirac systems scales as $\sqrt{\lvert n \lvert B}$, so that individual interband transitions can be addressed with high selectivity at a fixed photon energy simply by tuning \textit{B}. 
This provides precise control over which electronic transition contributes to the light-induced effective magnetic field. 
Moreover, because Landau quantization in Dirac systems is governed purely by the linear band dispersion rather than by details of the underlying lattice, they constitute a minimal model in which the electronic contribution to light-induced magnetism can be isolated from other possible sources, such as phonons or sublattice symmetry breaking.
Graphene, the prototypical Dirac material, exhibits pronounced Faraday rotation both in the classical Drude response and at resonant Landau-level transitions, as demonstrated by static magneto-optical spectroscopy across the far- and mid-infrared spectral
ranges~\cite{Crassee2011}.
This large magneto-optical susceptibility is indicative of a strong inverse Faraday effect and motivates the search for substantial light-induced magnetic fields in the time domain with microscopic understanding.
%large magneto-optical susceptibility indicates a significant inverse Faraday effect, potentially leading to substantial light-induced magnetic fields.

\begin{figure}[H]
  \centering
   \includegraphics[width=\linewidth]{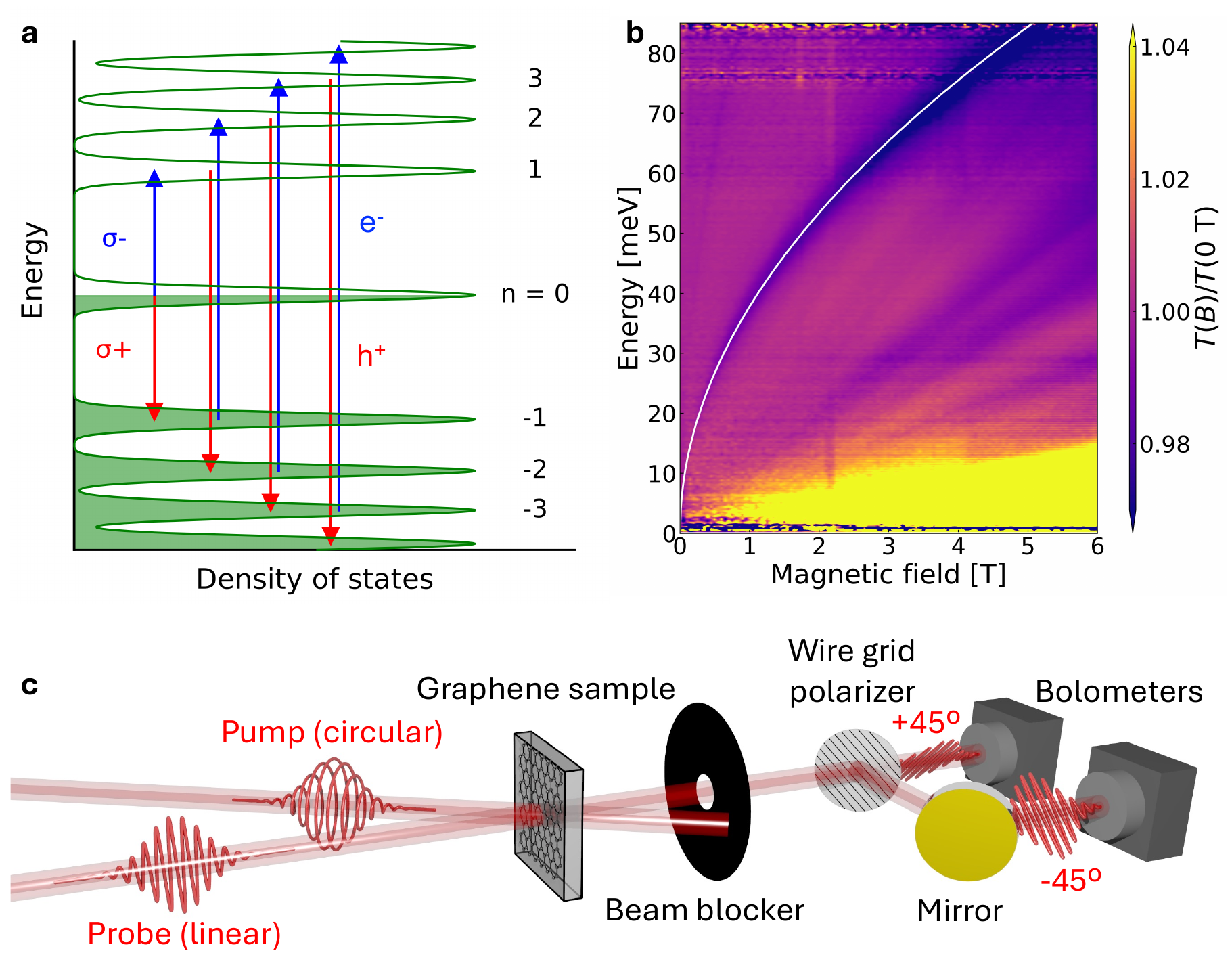}
  \caption{\textbf{Energy diagram and experimental setup.}
    \textbf{a},~Energy diagram of the Landau levels in graphene.
    Circular polarization selectively couples to electron (upward arrows) or hole (downward arrows) excitations between the levels.
    Red and blue arrows denote excitations by $\mathrm{\sigma}+$ and $\mathrm{\sigma}-$ polarizations, respectively.
    \textbf{b},~Magneto-transmission spectra of the measured graphene sample, where the white curve indicates the Landau-level transition from $n=0$ to $n=\pm1$, following $E(B)=\,v_\mathrm{F}\sqrt{2e\hbar B}$, as a guide to the eye.
    \textbf{c},~Schematic view of the time-resolved single-color experimental setup.
    The circular polarization of the pump beam is switched between $\sigma^+$ and $\sigma^-$ using a quarter waveplate.
    The non-collinear geometry allows blocking the pump beam by a beam blocker after the sample placed in a superconducting magnet, while the probe beam reaches the bolometers with balanced detection between the two arms with orthogonal polarizations ($\pm 45^\circ$).
    Details can be found in Methods.}
  \label{fig:1}
\end{figure}

Here, we report the first time-resolved observation of ultrashort effective magnetic fields in graphene generated by selective resonant excitation between the degenerate Landau-level transitions from $n = -1$ to $n = 0$ and from $n = 0$ to $n = +1$ using circularly polarized terahertz pulses (see Fig.~\ref{fig:1}\textbf{a}).
These degenerate transitions are known to produce qualitatively distinct carrier dynamics through complex Auger scattering involving equilibrium carriers in the $n = 0$ level~\cite{Mittendorff2014}.
We show that this selectivity also imprints on the time-resolved magneto-optical response.
By anchoring transient Faraday rotation signals to static magneto-optical spectra, we quantify the light-induced effective magnetic fields and find them to be on the order of 10 milli-Tesla (mT).
A magnetic field dependence of the transient signals across the resonance condition, exhibiting a dispersive structure and mirroring the static Faraday rotation lineshape, indicates that the pump-induced redistribution of Landau-level occupancies modifies the optical Hall conductivity and hence the Faraday rotation angle on femtosecond to picosecond timescales.
This constitutes a direct, time-resolved demonstration of a purely electronic inverse Faraday effect in a Dirac system, isolated from phonon and sublattice contributions by magnetic tuning of the Landau levels.

% ---- Figure 1 placeholder ----

\subsection*{Results}

\begin{figure}[H]
  \centering
   \includegraphics[width=\linewidth]{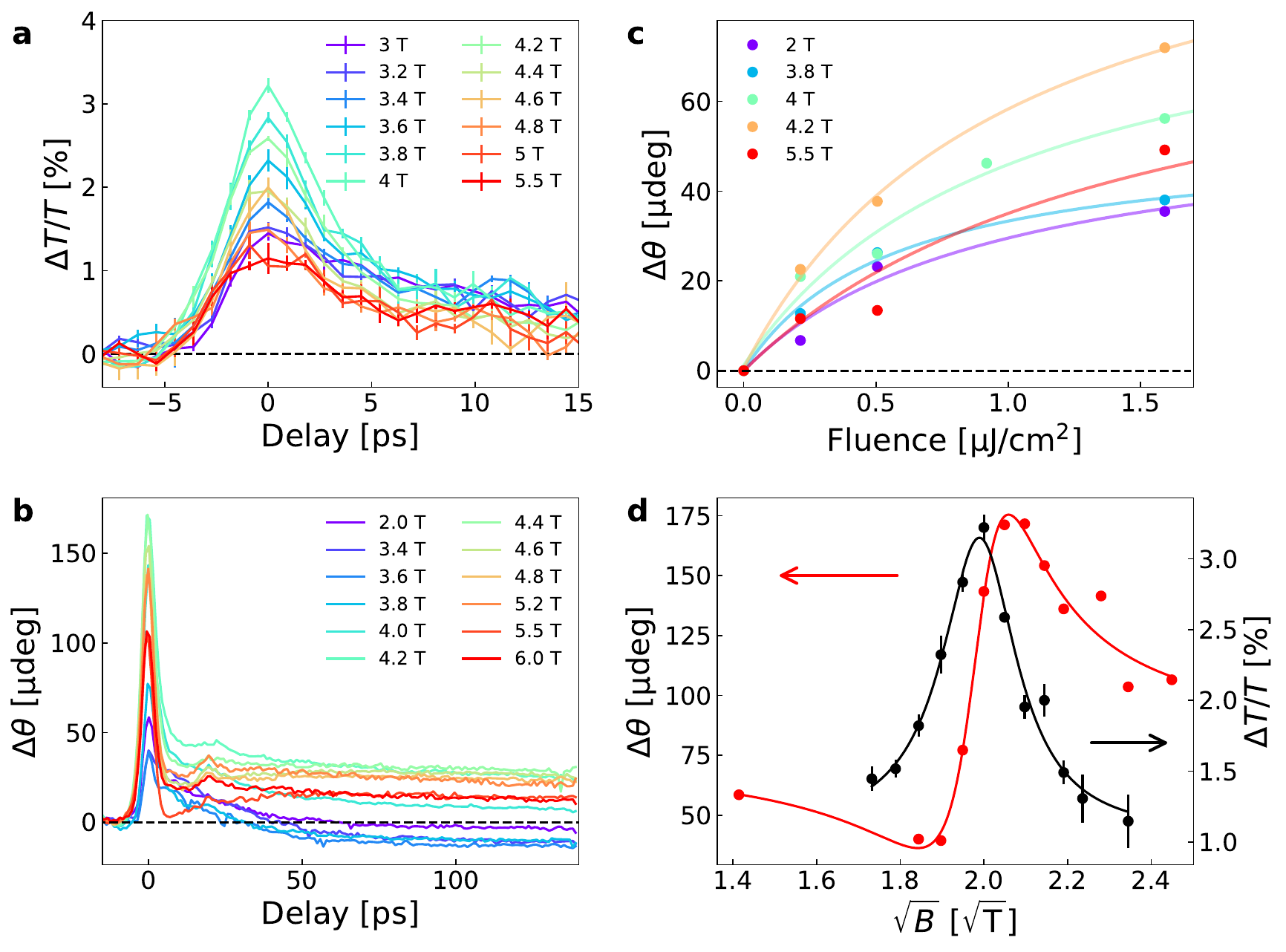}
  \caption{\textbf{Characterization of the dynamics triggered by circularly polarized terahertz pulses.}
    \textbf{a},~Differential transmission $\DeltaT/T$ and \textbf{b},~polarization rotation angle $\Deltatheta$ at different magnetic fields as a function of pump-probe delay.
    \textbf{c},~Fluence ($F$) dependence of peak $\Deltatheta$ signals at different magnetic fields.
    Curves are guides to the eye from the fit to the Langmuir saturation model, $\Delta\theta(F) = \Delta\theta_{\mathrm{max}} \frac{F}{F+F_{\mathrm{sat}}}$, where $\Delta\theta_{\mathrm{max}}$ and $F_{\mathrm{sat}}$ represent the saturation amplitude and the fluence over which saturation occurs, respectively.
    Obtained $F_{\mathrm{sat}}$ is $\sim1$ $\mathrm{\mu J/cm^{2}}$. 
    \textbf{d},~Magnetic field dependence of $\DeltaT/T$ (black) and $\Deltatheta$ (red), extracted from the data shown in \textbf{a} and \textbf{b}, respectively. 
    The black and red solid curves represent a fit to $\DeltaT/T$ by an asymmetric Lorentzian peak and to $\Deltatheta$ by an asymmetric Lorentzian dispersion, respectively.
    For all data, the circularly polarized pump beam helicity was $\sigma^+$ and the magnetic field was along the positive direction, $(\sigma^+,\,+B)$.
    Error bars in \textbf{a} and \textbf{d} represent the standard errors among multiple scans, while data in \textbf{b} and \textbf{c} were taken by single scans using a lock-in amplifier (see Methods).}
  \label{fig:2}
\end{figure}

Our experiments employ a single-color pump-probe scheme at a photon energy of \SI{75}{\meV} (see Fig.~\ref{fig:1}\textbf{c} and Methods for details).
The pump pulse was circularly polarized ($\sigma^+$ for right and $\sigma^-$ for left) to selectively excite the Landau-level transitions, while the probe pulse was linearly polarized. 
Its polarization rotation after the sample was analyzed using balanced detection.
The Landau-level energies were tuned by the applied field $B$, where ${\sim}\,\SI{4.2}{T}$ brings the first degenerate resonance (hole excitation from $n = 0$ to $n = -1$ and electron excitation from $n = 0$ to $n = +1$; see Fig.~\ref{fig:1}\textbf{a}) into coincidence with the \SI{75}{\meV} photon energy, as represented by the fan diagram of magneto-optical transmission spectra (Fig.~\ref{fig:1}\textbf{b}). 
Resonant excitation is further verified by a transient increase in transmission around this field amplitude, obtained from the sum of the two bolometer signals and attributable to reduced Pauli blocking (see Figs. \ref{fig:2}\textbf{a} and \ref{fig:2}\textbf{d}).

% Circular polarization is defined as $(E_x,\,E_y) = e^{i(kz-\omega t)} \tfrac{1}{\sqrt{2}}(1,\,\pm i)$, where the upper and lower signs correspond to circular right ($\sigma^+$) and left ($\sigma^-$) polarizations, respectively.
% We take the Cartesian coordinate system in which $z$ is the propagation direction of electromagnetic waves while $x$ and $y$ are perpendicular to $z$, with wavenumber $k$ and angular frequency $\omega$.
Our sample is decoupled multilayer graphene epitaxially grown on the C-terminated face of a SiC substrate~\cite{Berger2006, Sprinkle2009}, as used in the prior time-resolved study~\cite{Mittendorff2014}.
Due to charge transfer at the interface, the layers closest to the substrate are heavily doped.
While these layers contribute to near-infrared absorption~\cite{Sun2008, Winnerl2013}, the mid-infrared excitation used here interacts predominantly with charge carriers in the nearly intrinsic layers further away from the substrate.
These layers are slightly $n$-doped, which breaks the electron-hole symmetry, strictly speaking~\cite{Orlita2008}.
From fits to the magneto-transmission spectra, we extract the Fermi energy $\EF$ and Fermi velocity $\vF$ as ${\sim}$15 meV and ${\sim}10^6\,\si{\metre\per\second}$, respectively, consistent with Ref.~\cite{CastroNeto2009}.
The corresponding excess carrier density $n_e$ at $B = 0\,\si{T}$ is $n_e \approx 1.7\times10^{10}\,\si{\cm^{-2}}$, while the Landau-level degeneracy $D$ at the first resonance at \SI{75}{\meV} and $B = \SI{4.2}{T}$ is $D \approx 40.6\times10^{10}\,\si{\cm^{-2}}$, taking the spin and valley degeneracy into account.
The well-separated Landau levels by ${\sim}\,\SI{75}{\meV}$ at \SI{4.2}{T} suppress thermal carrier excitation across the levels at 10 K, giving static occupancies of $N_{0, n=0} = n_e/D + 0.5 \approx 0.54$, $N_{0, n=+1} = 0$, and $N_{0, n=-1} = 1$.

Figure~\ref{fig:2}\textbf{b} shows the time traces of the polarization rotation angles $\Deltatheta$ measured with the pump polarization $\sigma^+$ in $+B$, $(\sigma^+,\,+B)$.
The peak values exhibit a dispersive field dependence across the first resonance (\SI{4.2}{T}; see Fig.~\ref{fig:2}\textbf{d}), qualitatively mirroring the static Faraday rotation lineshape~\cite{Crassee2011}.
The dynamics are long-lived, consistent with the absence of phonon modes near the Landau-level energies that could serve as dissipation channels~\cite{Wendler2013, Wendler2017}. 
This identifies the excited electronic system as the primary origin of the observed signals.
At later timescales (10's of picoseconds), the signals develop additional structure, including a sign flip, reflecting the involvement of Landau levels beyond $n=0$ and $\pm1$ through Auger scattering~\cite{Mittendorff2014}. 
While a complete account of these longer-lived dynamics lies beyond the scope of the present work, we note that they are reasonably attributed to the Auger-mediated redistribution of Landau-level occupancies reported previously~\cite{Mittendorff2014} and do not affect the conclusions drawn from the signals at fast timescale. 
The fluence dependence shows a deviation from linear growth and saturation around 1 $\mu\si{J/cm^2}$ (see Fig.~\ref{fig:2}\textbf{c}), also suggestive of rapid intra-pulse Auger scattering~\cite{Mittendorff2014}.

% ---- Figure 2 placeholder ----

\begin{figure}[H]
  \centering
   \includegraphics[width=\linewidth]{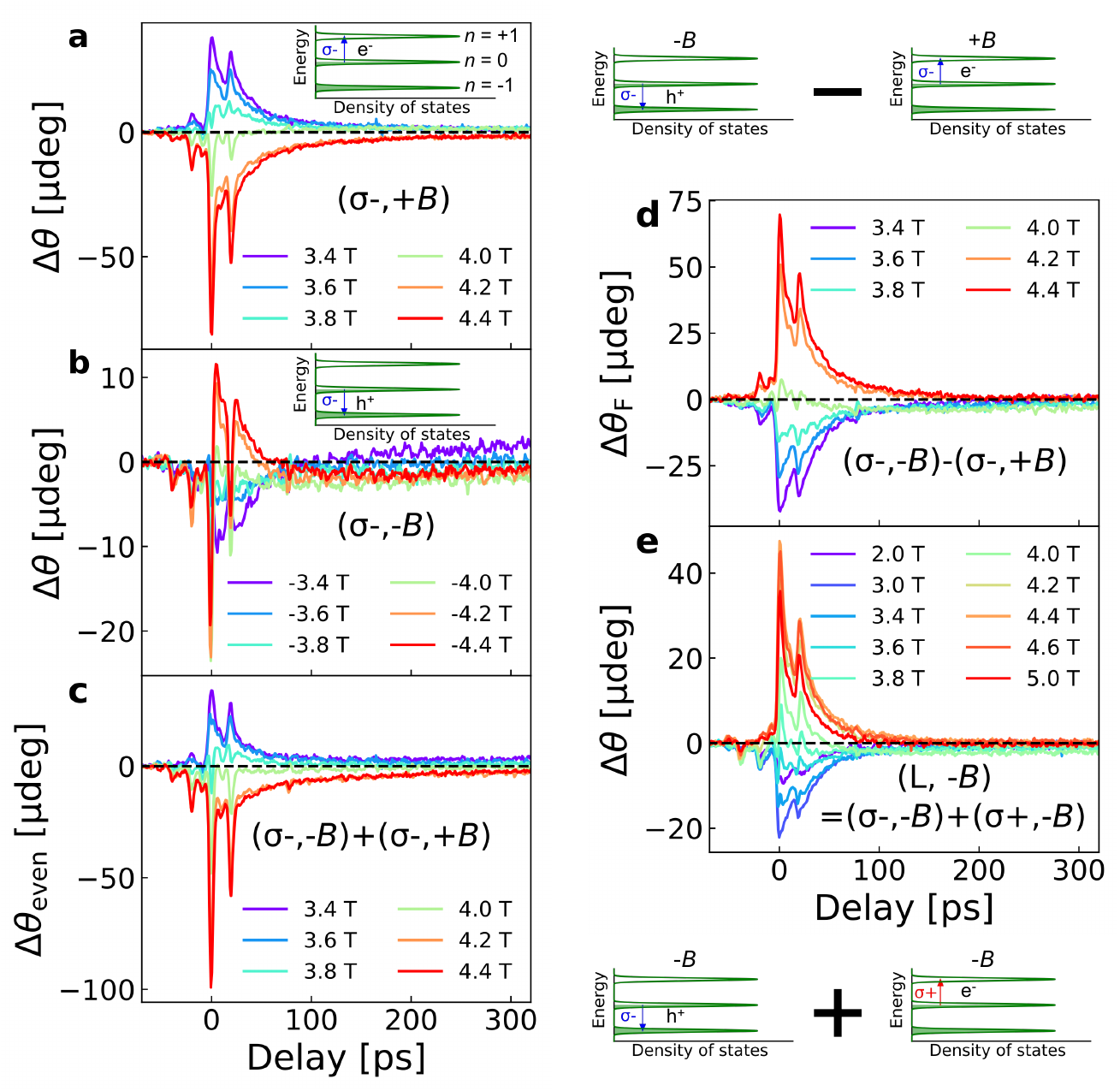}
  \caption{\textbf{Pump polarization and magnetic field dependences of the polarization rotation signals.}
    Polarization rotation angle $\Deltatheta$ measured at different magnetic fields and pump beam polarizations as a function of pump-probe delay: \textbf{a},~$(\sigma^-,\,+B)$, \textbf{b},~$(\sigma^-,\,-B)$, and \textbf{e},~$(L,\,-B)$.
    The inset in \textbf{a} and \textbf{b} represents the corresponding excitation diagram.
    \textbf{c},~Sum and \textbf{d},~difference of the data shown in \textbf{a} and \textbf{b}.
    The comparison between \textbf{d} and \textbf{e} demonstrates the equivalence between $(\sigma^+,\,-B)$ and sign-flipped $(\sigma^-,\,+B)$ data, as indicated by the excitation diagram shown above or below each panel (see main text for details).
    The difference data shown in \textbf{d} corresponds to the Faraday rotation $\DeltaF$, while the sum data shown in \textbf{c} corresponds to $B$-even or $B$-independent signals $\Delta\theta_{\mathrm{even}}$.}
  \label{fig:3}
\end{figure}

To confirm the involvement of magnetic (Faraday) signals in the observed polarization rotation, we compare two datasets acquired with the same pump polarization $\sigma^-$ in opposite polarities of $B$, $(\sigma^-,\pm B)$, shown in Figs.~\ref{fig:3}\textbf{a} and~\ref{fig:3}\textbf{b}.
The suppressed amplitude and sign flip shortly after the peak at $t=0$ ps observed for $(\sigma^-,-B)$ (Fig.~\ref{fig:3}\textbf{b}) are attributed to Auger scattering involving equilibrium carriers in the more than half-filled $n = 0$ level.
As observed previously in level-selective pump-probe experiments~\cite{Mittendorff2014}, this Auger scattering for slightly $n$-doped samples produces smaller amplitudes and sign flips between fast and slower components for the excitation between the levels ${n=-1}$ and ${n=0}$, while leaving no qualitative signatures of Auger scattering in the transition between the levels ${n=0}$ and ${n=+1}$.
We observe analogous behavior in the polarization rotation signals (compare Figs.~\ref{fig:3}\textbf{a} and~\ref{fig:3}\textbf{b}) and attribute it to the same mechanism.
The multiple-peak structure in some traces arises from internal reflections of the pump and probe beams within the SiC substrate.
This structure is less pronounced in the time traces shown in Fig.~\ref{fig:2}\textbf{b}, which we attribute to sensitivity of the signals to subtle changes in the optical path upon switching the pump helicity, as consistent with the non-collinear geometry, combined with spatial inhomogeneity across the sample.
Thus, even though the strong asymmetry between $\pm B$ already suggests a magnetic (Faraday) origin, careful isolation of the contribution is necessary.
We validate this using a symmetry argument detailed below.

Specifically, we compare the signals measured with circular polarization pump with those measured with linear polarization pump \textit{L}, which is the sum of opposite circular polarizations: $(L,\,-B) = (\sigma^-,\,-B) + (\sigma^+,\,-B)$.
The differential signals $(\sigma^-,\,-B) - (\sigma^-,\,+B)$ and their field evolution are qualitatively equivalent to the linear pump signals (compare Figs.~\ref{fig:3}\textbf{d} and~\ref{fig:3}\textbf{e}), indicating the equivalence between $(\sigma^+,\,-B)$ and sign-flipped $(\sigma^-,\,+B)$ data.
This equivalence has a transparent physical origin rooted in two ingredients.
(1) Landau levels in graphene carry pseudo-spin structures tied to orbital motions whose helicity selectivity is set by the polarity of $B$; reversing $B$ therefore exchanges the helicity selectivity of the transitions.
As a result, $(\sigma^+,\,-B)$ and $(\sigma^-,\,+B)$ are equivalent in driving the $n = 0 \to n = +1$ transition (see diagrams in Figs.~\ref{fig:3}\textbf{d} and~\ref{fig:3}\textbf{e}).
%The microscopic polarization-dependent transition process between the Landau levels and (2) their contributions to magneto-optical phenomena.
%(1) Landau levels in graphene have pseudo-spin structures tied to orbital motions, defined by the $B$ polarity, and the helicity selectivity arises due to the angular momentum of circularly polarized photons and the orbital motions.
%Consequently, reversing the direction of $B$ exchanges the helicity selectivity of the transitions.
%As a result, for example, $(\sigma^+,\,-B)$ and $(\sigma^-,\,+B)$ are equivalent in exciting carriers between the Landau levels from $n = 0$ to $n = +1$ (see the diagrams shown above or below Figs.~\ref{fig:3}\textbf{d} and~\ref{fig:3}\textbf{e}).
(2) The Faraday rotation angle $\thetaF(\omega)$ is proportional to the real part of the optical Hall conductivity $\sigmaxy(\omega)$~\cite{Morimoto2009}, which is proportional to the difference of the optical conductivity between circular polarization bases:
\begin{equation}
  \thetaF(\omega) \propto \mathrm{Re}\,\sigmaxy(\omega)
  \propto \mathrm{Re}\bigl[\sigma^+(\omega) - \sigma^-(\omega)\bigr].
  \label{eq:1}
\end{equation}
Reversing $B$ swaps the helicity assignments for a given transition, inverting the sign of its Faraday contribution.
%Therefore, flipping the $B$ polarity, which swaps the circular polarization helicity for a transition between specific Landau levels, inverts the sign of Faraday rotation angles associated with a specific Landau level transition.
Together, (1) and (2) imply that data with $(\sigma^+,\,-B)$ and sign-flipped data with $(\sigma^-,\,+B)$ are equivalent for $\thetaF$.
The experimental observation of this equivalence validates the internal consistency of the $(\sigma^-,\pm B)$ dataset and the involvement of the Faraday signals in the observed time-dependent response.

We isolate the magnetic (Faraday) contribution by taking the $B$-asymmetric combination:
\begin{equation}
  \DeltaF = (\sigma^-,\,-B) - (\sigma^-,\,+B),
  \label{eq:2}
\end{equation}
plotted in Fig.~\ref{fig:3}\textbf{d}.
This quantity is, by construction, independent of any $B$-even or $B$-independent effects, such as birefringence and magnetostriction.
The complementary $B$-symmetric combination 
\begin{equation}
  \Delta\theta_{\mathrm{even}} = (\sigma^-,\,+B) + (\sigma^-,\,-B),
  \label{eq:3}
\end{equation}
shown in Fig.~\ref{fig:3}\textbf{c}, captures these contributions.
While $\DeltaF$ time traces display a relatively simple temporal profile, with a fast rise and an exponential decay, apart from multiple peaks due to internal reflections, $\Delta\theta_{\mathrm{even}}$ time traces show a more complex structure comprising a short-lived peak and an independent slow decay.
The short-lived peak likely reflects an initial non-equilibrium excitation of the electronic system during the pulse, which transiently modifies the linear birefringence contribution.
The slow decay indicates a thermalized hot-electron contribution with a long relaxation time constant due to the suppressed electron-phonon coupling near the excitation energy, around which phonon modes are absent~\cite{Wendler2013, Wendler2017}, in addition to the involved higher Landau levels through Auger scattering.
These contrasting behaviors confirm that $\DeltaF$ isolates the intrinsic Faraday response.

% ---- Figure 3 placeholder ----

The $B$ evolution of $\DeltaF$ time traces (see Fig.~\ref{fig:3}\textbf{d}) reveals a dispersive structure with the signal changing sign across the resonant field.
The lineshape qualitatively mirrors the static Faraday rotation spectra (see Figs.~\ref{fig:4}\textbf{a} and~\ref{fig:4}\textbf{b}), proportional to $\mathrm{Re}[\sigma^+(\omega) - \sigma^-(\omega)]$ [see Eq.~\eqref{eq:1}], and indicates that selective resonant excitation between the Landau levels generates an additional magnetic field on top of the applied magnetic field $B$.
%Here, the two-step structure visible in the static data reflects the slight breaking of electron-hole degeneracy due to the finite $n$-doping, which shifts the resonances from $n=-1$ to $n=0$ and from $n=0$ to $n=+1$ by a small but resolvable amount.
%Since the circularly polarized pump beam selectively drives one of the Landau level transitions, the sign of $\DeltaF$ is intact once it goes across the first resonant magnetic field.
Quantitative comparison between the transient and static Faraday rotation signals yields the time traces of induced magnetic fields, shown in Fig.~\ref{fig:4}\textbf{c}.
%Due to the dispersion of $\DeltaF$ across the zero-point, the effective magnetic field has a node around the resonant magnetic field (see Fig.~\ref{fig:4}\textbf{b}). 
The induced field reaches values on the order of magnitude of 10 mT at the fluences employed here around 1.6 $\mu\si{J/cm^2}$, corresponding to a substantially smaller electric field of $\sim$20 kV/cm, compared to typical values on the order of 1--10 MV/cm generating large light-induced magnetic fields~\cite{Kirilyuk2010, Stanciu2007, Kimel2005, Sheu2019, Kozhaev2018, Hansteen2005, Tzschaschel2019}.
The effective magnetic field normalized by the pump electric field $B_{\mathrm{eff}}/E_{\mathrm{pump}}$ is summarized in Table~\ref{tab:table1} with values previously reported through the inverse Faraday effect of electronic origin.
The largest value of this work indicates the high efficiency of the inverse Faraday effect via the selective Landau-level transition mechanism.
%Due to similar lineshapes between $\DeltaF$ and static Faraday signals, the induced peak magnetic field is nearly constant across the magnetic field range.

\begin{table}[h]
    \centering
    \caption{Effective magnetic fields and pump electric fields via the inverse Faraday effect of electronic origin.}
    \label{tab:table1}
    \begin{tabular}{|c|c|c|c|c|}
        \hline
        Wavelength [$\mathrm\mu$m] & $B_{\mathrm{eff}}$ [mT] & $E_{\mathrm{pump}}$ [MV/cm] & $B_{\mathrm{eff}}/E_{\mathrm{pump}}$ [mT cm/MV] & Reference\\
        \hline
        0.400 & 100 & 0.27 & 370 & \cite{Kirilyuk2010}\\
        0.642 & 84 & 1.76 & 47.7 & \cite{Kozhaev2018} \\
        0.800 & 10 & 4.2 & 2.3 & \cite{Stanciu2007} \\
        0.800 & 300 & 10.3 & 29 & \cite{Kimel2005} \\
        0.805 & 600 & 8.4 & 71 & \cite{Hansteen2005} \\
        1.240 & 300 & 1.88 & 160 & \cite{Sheu2019} \\
        1.278 & 760 & 18.1 & 42 & \cite{Tzschaschel2019} \\
%        28.3 $\mathrm{\mu}$m & 920 & 0.5 & 1840 & \cite{Luo2023} \\
%        99.9 $\mathrm{\mu}$m & 32 & 0.23 & 140 & \cite{Basini2024} \\
        16.5 & 10 & 0.02 & 500 & this work \\
        \hline
    \end{tabular}
\end{table}

\subsection*{Discussion}

%Selective optical excitation between Landau levels can, in principle, generate transient magnetic fields through two distinct mechanisms.
%(1) The first is a real orbital mechanism: different Landau levels carry orbital magnetizations, so that a pump-induced redistribution generates a real magnetic field.
%(2) The second is an optical Hall mechanism: different Landau levels contribute differently to $\sigmaxy(\omega)$ and corresponding Faraday rotations, represented by effective magnetic fields.
%These two mechanisms produce qualitatively distinct transient Faraday rotation lineshapes.
%For the real orbital mechanism, this would produce a signal proportional to \frac{\partial\thetaF}{\partial B}\, i.e., the derivative of a dispersive lineshape, which is an symmetric peaks.
%In contrast

Selective optical excitation between Landau levels can, in principle, generate transient magnetic fields through two distinct mechanisms, corresponding to two different physical origins of the magneto-optical response.
The first is a real orbital mechanism: different Landau levels carry different orbital magnetization, so that a pump-induced redistribution of level occupancies alters the macroscopic magnetization arising from diamagnetism~\cite{Vallejo2021} and thereby generates a real transient magnetic field.
The second is an optical Hall mechanism: different Landau levels contribute differently to $\sigmaxy(\omega)$ and hence to the Faraday rotation, so that a redistribution of occupancies via the selective optical excitation directly modifies $\sigmaxy(\omega)$, represented by an effective field.

\begin{figure}[H]
  \centering
   \includegraphics[width=\linewidth]{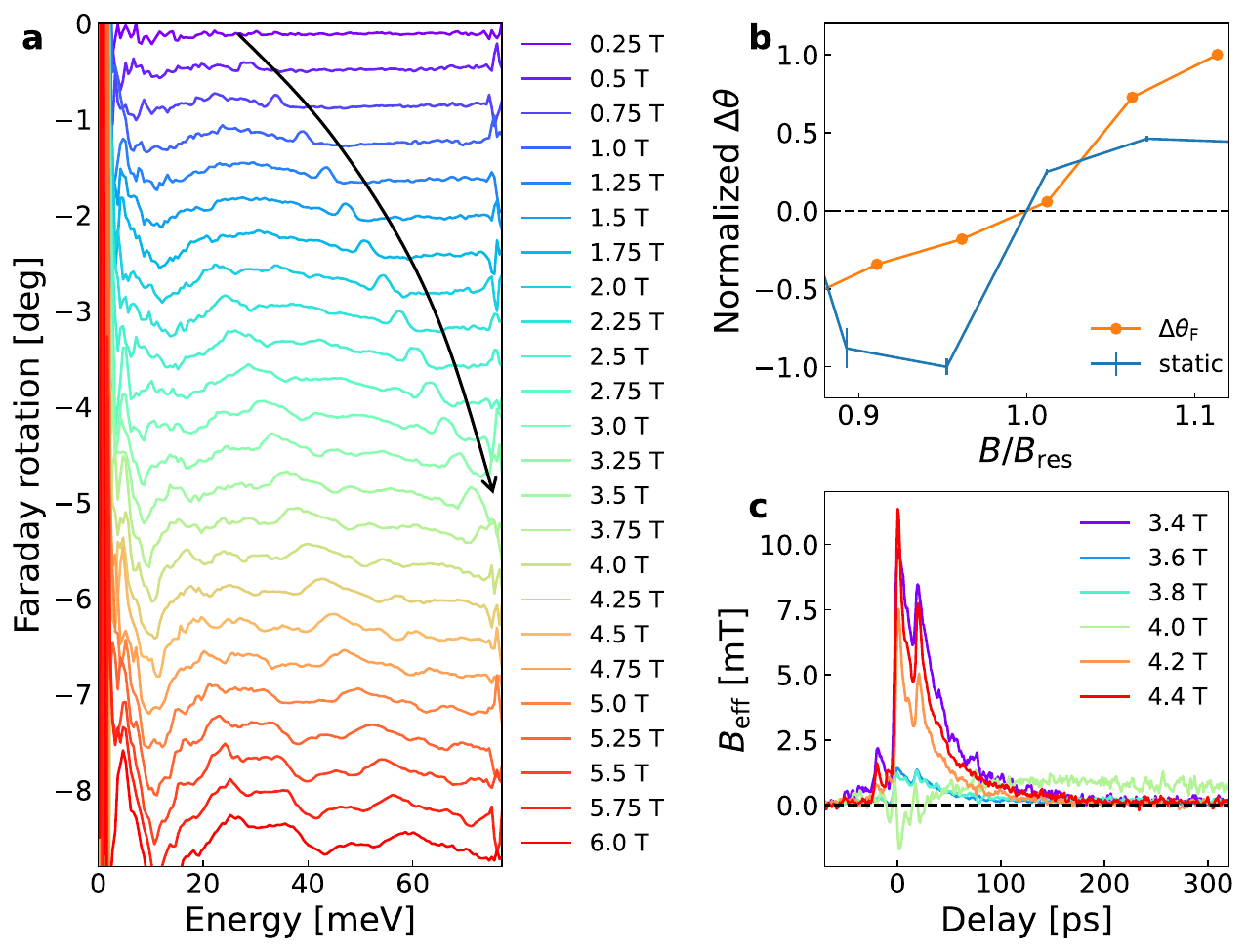}
  \caption{\textbf{Effective magnetic fields extracted by comparison to static Faraday rotation spectra.}
    \textbf{a},~Static Faraday rotation spectra measured at different magnetic fields.
    Each data is vertically shifted for clarity.
    The black arrow is a guide to the eye to follow Faraday rotation signals due to the first Landau-level transition between $n=0$ and $n=\pm1$.
    \textbf{b},~$B$ evolution of the static and transient Faraday rotation signals (blue and orange, respectively) normalized for comparison.
    Because of instrumental artifacts near 75 meV in the static data, the data around 71 meV is shown instead, normalized to the resonant field.
    Here, $\DeltaF$ represents the peak values in the time traces.
    \textbf{c},~Time traces of the effective magnetic field ($\Beff$) at different applied magnetic fields, obtained by comparing the static and transient Faraday rotation signals.}
  \label{fig:4}
\end{figure}

These two mechanisms produce qualitatively distinct transient Faraday rotation lineshapes.
For the real orbital mechanism, the induced Faraday rotation is proportional to the derivative of $\DeltaF$ with respect to $B$, since the transient field is much smaller than the applied field:
%Given the different orders of the field strength between the transient magnetic field and the applied magnetic field, if the observed signals are due to real magnetic fields, the induced Faraday rotation angles are proportional to the derivative with respect to $B$:
\begin{equation}
  \DeltaF = \thetaF(\omega,\,B+\Delta B) - \thetaF(\omega,\,B)
  \approx \frac{\partial\thetaF}{\partial B}\,\Delta B.
  \label{eq:4}
\end{equation}
In contrast, the optical Hall mechanism modifies $\sigmaxy(\omega)$ directly, producing a signal that mirrors the dispersive static Faraday rotation lineshape.
The experimentally observed dispersive $B$-dependence of $\DeltaF$ (see Fig.~\ref{fig:4}\textbf{b}) is inconsistent with the derivative lineshape expected from the real orbital mechanism and instead matches the optical Hall prediction, identifying the latter as the dominant contribution.
%The observed dispersive structure, similar to the static Faraday rotation signals, excludes the first mechanism.

Within the optical Hall mechanism, the change in $\sigmaxy(\omega)$ due to pump-induced occupancy redistribution among the three relevant Landau levels is~\cite{Gusynin2007}:
%Direct excitation between the Landau levels can modify $\sigmaxy(\omega)$.
%In the case of occupancy changes among the three Landau levels, the change in $\sigmaxy(\omega)$ is expressed as~\cite{Gusynin2007}:
\begin{equation}
  \Deltasigmaxy(\omega) \propto
  \Delta\bigl(2N_{n=0} - N_{n=+1} - N_{n=-1}\bigr).
  \label{eq:5}
\end{equation}
The measured transient Faraday rotation is approximately 50 times smaller than the equilibrium value.
Together with the carrier number conservation law ($\Delta N_{n=+1} + \Delta N_{n=-1} = -2\Delta N_{n=0}$), Eq.~\eqref{eq:5} implies that the observed signals correspond to a change of $N_{n=0}$ by ${\sim}\,1\%$.
This small occupancy change is consistent with the relatively low fluences and corresponding electric fields used in our experiment ($\sim$1.6 $\mu\si{J/cm^2}$ and $\sim$20 kV/cm) and with rapid intra-pulse Auger scattering that partially re-equilibrates the carriers within the pulse duration~\cite{Mittendorff2014}.
%In the prior study~\cite{Mittendorff2014}, occupancy changes exceeding $5\%$ were reported for a multilayer graphene sample with a similar $n_e$ in the order of $10^{10}$ cm$^{-2}$, which, through Eq.~\eqref{eq:5}, would produce $\sigmaxy$ changes above 10\% and corresponding effective magnetic fields on the order of $10^{-1}$ T.
%The mT-scale fields reported here thus reflect the low-fluence, low-doping regime of the present experiment, and substantially larger effective fields are achievable by increasing fluence or by modifying the carrier density toward charge neutrality.

%leading to $>10\%$ changes in $\sigmaxy(\omega)$.
%Corresponding effective magnetic fields reach on the order of $10^{-1}$ T.
%Different doping levels and the number of intrinsic graphene layers could significantly enhance the effective magnetic fields.

These results establish a direct, time-resolved link between resonant Landau-level excitation and modification of the optical Hall conductivity in a Dirac system.
The ability to selectively drive the Landau-level transition and hence transiently modulate $\sigmaxy(\omega)$ opens the possibility of using circularly polarized terahertz pulses as a tool to manipulate magneto-optical responses on ultrafast timescales in systems with Landau quantization.
The formalism demonstrated here provides a clean model system of the inverse Faraday effect with an electronic origin, yielding effective magnetic fields.
Electrostatic tunability of equilibrium occupancies by gate voltage~\cite{Zhang2005} may provide a new dimension to control the observed light-induced effective magnetic fields.
% ---- Figure 4 placeholder ----

In conclusion, we have demonstrated the emergence of ultrashort effective magnetic fields in graphene through selective resonant excitation between non-equidistant Landau quantized levels arising from the linear Dirac dispersion of electronic bands.
The dispersive transient Faraday rotation lineshape identifies the optical Hall conductivity, rather than real orbital magnetization, as the origin of the light-induced effective field.
Magnetic tuning of the Landau-level energies provides energetic isolation of the electronic excitation from any sublattice contributions, such as phonons and plasmons, making graphene a clean and ideal platform for the demonstration.
%Energetic isolation of the electronic excitation from other possible excitations by magnetic tuning of the Landau energy levels makes it a clean platform for demonstrating the emergent effective magnetic fields of purely electronic origin.
The effective magnetic field experienced by the electronic system is transiently modified through the optical excitation, with amplitudes reaching the 10 mT scale, which is, when normalized to the pump electric fields, larger than other studies reporting light-induced magnetic fields via the inverse Faraday effect of electronic origin~\cite{Kirilyuk2010, Stanciu2007, Kimel2005, Sheu2019, Kozhaev2018, Hansteen2005, Tzschaschel2019}.
%require scaling by 2-3 orders of magnitude for comparison to the other studies reporting light-induced magnetic fields~\cite{Kirilyuk2010, Stanciu2007, Kimel2005, Satoh2010, Sheu2019, Nemec2018, Basini2024}, considering the used electric field amplitude difference.
By adjusting the applied field to match the energy scale of the Landau-level transitions to those of other low-energy excitations, e.g., phonons, magnons, polarons, plasmons, or excitons, whether such transient modifications of the electronic Landau-level occupancy or emergent ultrafast effective magnetic fields subsequently drive the other low-energy excitations through electron-phonon, electron-magnon, or related coupling mechanisms is a question of fundamental interest.
%Whether tuning of the Landau levels subsequently affects other low-energy excitations, e.g., phonons, magnons, polarons, and excitons, via electron-phonon, electron-magnon, or other coupling mechanisms is a question of fundamental interest.
For example, a substantially enhanced phonon magnetic moment has been reported when the cyclotron resonance is tuned into coincidence with a degenerate phonon mode~\cite{Cheng2020b}, valley-selective optical excitation using circularly polarized optical pulses shifts exciton energies~\cite{Sie2014}, and recent theoretical work suggests that electronic band splitting may play a central role in phonon magnetism~\cite{Klebl2025}.
Dirac systems potentially host axion electrodynamics at the surface, manifested by quantized magneto-optical effects~\cite{Wu2016}, and dynamical tuning of carrier densities might transiently affect such exotic physics via the effective magnetic fields.
Our demonstration forms a foundation for exploring emergent magneto-optical phenomena in Dirac systems and related materials via ultrafast effective magnetic fields, with potential for ultrafast control of spin, valley, lattice, and magneto-electric functionalities.
% In principle, such an excitation can give rise to real magnetic fields due to direct modulation of the number of carriers exhibiting orbital motion.
% Regardless, the observation is consistently explained by effective magnetic fields, as indicated by changes in the optical Hall conductivity, which result in transient magnetic signals as if there were a modulation of the applied magnetic field.
%\hl{Some future perspective on light-induced magnetic field effects on Dirac physics?}
% The formation of Landau levels is universal across many classes of materials once a magnetic field is applied.

% ============================================================
\section*{Methods}
% ============================================================
\textbf{Single-color pump-probe experiment}:
The single-color pump-probe experiment was performed at FELBE [free-electron lasers at the ELBE (electron linear accelerator with high brilliance and low emittance)] in Helmholtz-Zentrum Dresden-Rossendorf. The experimental setup is shown in Fig.~\ref{fig:1}\textbf{c}.
The central energy and the bandwidth of the pulses were 75 meV and 1.2 meV (1.6$\%$), respectively. 
The repetition rate was 13 MHz, and the pulse duration was $\sim$2.7 ps in full width at half-maximum.
The pump pulse was circularly polarized ($\sigma^+$ for right and $\sigma^-$ for left) to selectively excite the Landau-level transitions.
The helicity of the circular polarization pulses is defined as $(E_x,\,E_y) = e^{i(kz-\omega t)} \tfrac{1}{\sqrt{2}}(1,\,\pm i)$, where the upper (lower) sign represents circular-right (left) polarization, controlled by the tunable waveplate (PO-TWP-L4-25-FIR, ALPHALAS).
The maximum averaged beam power was about 60 mW with the quarter waveplate and a chopper.
The corresponding maximum pump laser fluence and electric field were $\sim$1.6 $\mu\si{J/cm^2}$ and $\sim$20 kV/cm, respectively.
The probe pulses were linearly polarized from the FELBE.

Our sample was cooled down to 10 K and placed in a superconducting magnet (Oxford Instruments).
Here, the positive (negative) magnetic field direction is defined as antiparallel (parallel) to the beam.
The pump and probe beams are non-collinear, allowing us to block the pump beam after the sample.
The probe beam is split into two paths after the sample for orthogonal polarizations ($\pm 45^\circ$) by using a wire grid polarizer (2-BBMP-H4-25, Altechna).
The azimuthal angle of the wire grid polarizer was adjusted to obtain similar intensities between the two paths for balanced detection.
The beam intensities were measured by two identical silicon bolometers (IRLabs Inc).
Differential signals between the two paths were collected by using a lockin-amplifier (SR830, Stanford Research Systems).
Noise levels were 0.3$\mu\mathrm{deg}$ in standard deviation.
A transmission change was measured by summing up the intensities of the two paths, which cancel the magnetic (Faraday) effect.

We used the same decoupled multilayer graphene sample for the pump-probe and static measurements. 
It was epitaxially grown on the C-terminated face of SiC, leading to highly doped layers near the substrate and slightly \textit{n}-doped layers for the remained part. 
The Fermi energy $\EF$ is estimated as 15 meV by extrapolating the Landau-fan diagram shown in Fig~\ref{fig:1}\textbf{b}.
The Fermi velocity $\vF$ of graphene is estimated as ${\sim}10^6\,\si{\metre\per\second}$ from the white curve shown in Fig.~\ref{fig:1}\textbf{b}, indicating the first Landau-level transition.
The value is consistent with Ref.~\cite{CastroNeto2009}.
While the majority of layers are uncoupled monolayers, there are some bilayer and multilayer contributions, as indicated by the Landau-level transitions at lower energy with almost linear energy dependence (see Fig.~\ref{fig:1}\textbf{b})~\cite{Orlita2011}.
As these lines do not cross the monolayer  $n=0$ to $n=+1$ resonance around 75 meV, we assume that they do not contribute to the Faraday rotation in our experiment.
The excess carrier density $n_e$ at $B = 0\,\si{T}$ is
\begin{equation}
  n_e = \frac{\EF^2}{\pi\hbar^2\vF^2} = 1.7\times10^{10}\,\si{\cm^{-2}},
\end{equation}
while the Landau-level degeneracy $D$ at the first resonance at \SI{75}{\meV} and $B = \SI{4.2}{T}$ is
\begin{equation}
  D = \frac{4eB}{h} = 40.6\times10^{10}\,\si{\cm^{-2}},
\end{equation}
 where 4 represents the spin and valley degeneracy.

\textbf{Static magneto-optical and transmission spectroscopy measurements}:
To collect the magneto-transmission data in Fig.~\ref{fig:1}\textbf{b}, a macroscopic area of the sample of about 4 mm$^2$ was exposed to the radiation of a globar or mercury lamp, which was analyzed by a Fourier transform spectrometer, and using light-pipe optics delivered to the sample placed in a superconducting coil.
The transmitted light was detected by a composite bolometer placed directly below the sample, both kept in low-pressure helium exchange gas at a temperature of 4.2 K.
The sample transmission $T(B)$  at a given magnetic field $B$ was normalized by the transmission data at $B=0$ to follow the $B$-dependent features only.
For Faraday rotation measurements presented in Fig.~\ref{fig:4}\textbf{a}, the sample was surrounded by two linear polarizers with axes mutually rotated by the angle of $\pm$45 deg.
These polarizers were made of a metallic grid defined holographically on a polyethylene substrate (TYDEX).
The Faraday angle $\theta_{\mathrm{F}}$ was then calculated using the formula
\begin{equation}
  \theta_{\mathrm{F}} = \pm \frac{1}{2}\frac{S(B_+)-S(B_-)}{S(B_+)+S(B_-)}.
  \label{eq:8}
\end{equation}
 where $S$ stands for single-channel spectrum recorded at the positive or negative magnetic field $B$.
This formula is valid at small angles that are relevant to our experiment. 

% ============================================================
\section*{Acknowledgments}
% ============================================================

The pump-probe experiment was performed at ELBE in the Helmholtz-Zentrum Dresden-Rossendorf e.\,V., a member of the Helmholtz Association (proposal No. 25203822-ST).
Authors acknowledge the support of the LNCMICNRS, a member of the European Magnetic Field Laboratory (EMFL).
We thank J. Budic for experimental assistance.

% ============================================================
\section*{Competing Interests}
% ============================================================

The authors declare no competing interests.

% ============================================================
\section*{Data Availability}
% ============================================================

Experimental and model data are accessible from the PSI Public Data Repository~\cite{DataRepository}.

% ============================================================
\section*{Author contributions}
% ============================================================

HU and SW conceived and designed the project.
JK prepared the graphene sample.
HU, AP, EU, MH, and SW discussed and performed the single-color pump-probe experiments.
KK, FC, and MO measured the static magneto-transmission and Faraday rotation spectra.
HU analyzed the experimental data.
HU and SW interpreted the results and wrote the manuscript with contributions from all authors.

% ============================================================
%  References  (compiled from references.bib)
% ============================================================
%\bibliographystyle{naturemag}   % naturemag.bst is available on Overleaf
\bibliography{references}       % points to references.bib

\end{document}